\documentclass[aps,prd,
letterpaper]{revtex4}

\usepackage[left=2cm, right=2cm, top=25mm, bottom=20mm]{geometry}
\usepackage{layout}
\usepackage{amsmath,amssymb,epsfig}
\usepackage{lscape}
\usepackage{color}
\usepackage{graphicx}
\usepackage{url}
\usepackage{comment}
\usepackage{lineno}
\usepackage{multirow}
\usepackage{rotating}
\usepackage{array}

\newcolumntype{x}[1]{%
>{\centering}m{#1}}%

\newcommand\etal{{\em et al.}}
\newcommand\dbd{\ensuremath{0\nu\beta\beta}}

\begin{document}

\title{{\bf {\huge R\&D towards CUPID}} \\ ( {\bf C}UORE {\bf U}pgrade with {\bf P}article {\bf ID}entification)}


\author{G.~Wang}
\author{C.L.~Chang}
\author{V.~Yefremenko}
\affiliation{High Energy Physics Division, Argonne National Laboratory, Argonne, IL, USA}

\author{J.~Ding}
\author{V.~Novosad}
\affiliation{Materials Science Division, Argonne National Laboratory, Argonne, IL, USA}

\author{C.~Bucci}
\author{L.~Canonica}
\author{P.~Gorla}
\author{S.S.~Nagorny}
\author{C.~Pagliarone}
\altaffiliation{Also with: University of Cassino,  Cassino Frosinone, Italy}
\author{L.~Pattavina}
\author{S.~Pirro}
\author{K.~Schaeffner}
\affiliation{INFN - Laboratori Nazionali del Gran Sasso, Assergi (AQ), Italy}

\author{J.~Feintzeig}
\author{B.K.~Fujikawa}
\author{Y.~Mei}
\affiliation{Nuclear Science Division, Lawrence Berkeley National Laboratory, Berkeley, CA, USA}

\author{E.B.~Norman}
\author{B.S.~Wang}
\affiliation{Department of Nuclear Engineering, University of California, Berkeley, CA, USA}

\author{T.I.~Banks}
\author{Yu.G.~Kolomensky}
\altaffiliation{Also with: Physics Division, Lawrence Berkeley National Laboratory, Berkeley, CA, USA}
\author{R.~Hennings-Yeomans}
\author{T.M.~O'Donnell}
\author{V.~Singh}
\affiliation{Department of Physics, University of California, Berkeley, USA}

\author{N.~Moggi}
\author{S.~Zucchelli}
\affiliation{Universit\`a di Bologna and INFN Bologna, Bologna, Italy}

\author{L.~Gladstone}
\author{L.~Winslow}
\affiliation{Massachusetts Institute of Technology, Cambridge, MA, USA}

\author{D.R.~Artusa, F.T.~Avignone III, R.J.~Creswick, H.A.~Farach, C.~Rosenfeld, J.~Wilson}
\affiliation{Department of Physics and Astronomy, University of South Carolina, Columbia, SC, USA}

\author{J.~Lanfranchi, S.~Sch\"onert, M.~Willers}
\affiliation{Technische Universit\"at M\"unchen, Physik-Department E15, Garching, Germany}

\author{S.~Di Domizio, M.~Pallavicini}
\affiliation{Dipartimento di Fisica, Universit\`a di Genova and INFN - Sezione di Genova, Genova, Italy}

\author{M.~Calvo, A.~Monfardini}
\affiliation{Institut N\'eel,CNRS and Universit\'e de Grenoble, INP, Grenoble,
France}

\author{C.~Enss, A.~Fleischmann, L.~Gastaldo}
\affiliation{Kirchhoff-Institute for Physics, University of Heidelberg,
Heidelberg, Germany} 

\author{R.S.~Boiko, F.A.~Danevich, V.V.~Kobychev}
\author{D.V.~Poda}
\altaffiliation{Also with: Centre de Sciences Nucleaires et de
  Sciences de la Matiere (CSNSM), CNRS/IN2P3, Orsay, France}
\author{O.G.~Polischuk}
\altaffiliation{Also with: INFN, Sezione di Roma ``La Sapienza'', Rome, Italy}
\author{V.I.~Tretyak}
\altaffiliation{Also with: INFN, Sezione di Roma ``La Sapienza'', Rome, Italy}
\affiliation{Institute for Nuclear Research, Kyiv, Ukraine}

\author{G.~Keppel, V.~Palmieri}
\affiliation{INFN - Laboratori Nazionali di Legnaro, Legnaro, Italy}

\author{K.~Kazkaz, S.~Sangiorgio, N.~Scielzo}
\affiliation{Lawrence Livermore National Laboratory, Livermore, CA, USA}

\author{K.~Hickerson, H.~Huang}
\affiliation{Department of Physics and Astronomy, University of California, Los Angeles, CA, USA}

\author{M.~Biassoni, C.~Brofferio, S.~Capelli, D.~Chiesa, M.~Clemenza,
O.~Cremonesi, M.~Faverzani, E.~Ferri, E.~Fiorini, A.~Giachero,
L.~Gironi, C.~Gotti, A.~Nucciotti, M.~Pavan, G.~Pessina,
E.~Previtali, C.~Rusconi, M.~Sisti, F.~Terranova}
\affiliation{INFN sez.~di Milano Bicocca and 
Dipartimento di Fisica, Universit\`a di Milano Bicocca, Milano, Italy}

\author{A.S.~Barabash, S.I.~Konovalov, V.V.~Nogovizin, V.I.~Yumatov}
\affiliation{State Scientific Center of the Russian Federation - Institute of Theoretical and Experimental Physics (ITEP), Moscow, Russia}

\author{F.~Petricca, F.~Pr\"obst, W.~Seidel}
\affiliation{Max-Planck-Institut f\"ur Physik, D-80805 M\"unchen, Germany}

\author{K.~Han, K.M.~Heeger, R.~Maruyama, K.~Lim}
\affiliation{Wright Laboratory, Department of Physics, Yale University, New Haven, CT, USA}

\author{N.V.~Ivannikova, P.V.~Kasimkin, E.P.~Makarov, V.A.~Moskovskih,
V.N.~Shlegel, Ya.V.~Vasiliev, V.N.~Zdankov}
\affiliation{Nikolaev Institute of Inorganic Chemistry, SB RAS, Novosibirsk, Russia}

\author{A.E.~Kokh, V.S.~Shevchenko, T.B.~Bekker}
\affiliation{Sobolev Institute of Geology and Mineralogy, SB RAS, Novosibirsk, Russia}

\author{A.~Giuliani, P.~de Marcillac, S.~Marnieros, E.~Olivieri}
\affiliation{Centre de Sciences Nucl\`eaires et de Sciences de la Mati\`ere (CSNSM), CNRS/IN2P3, Orsay, France}

\author{L.~Taffarello}
\affiliation{INFN - Sezione di Padova, Padova, Italy}

\author{M.~Velazquez}
\affiliation{Institut de Chimie de la Mati\`ere Condens\'e de Bordeaux (ICMCB), CNRS, 87, Pessac, France}

\author{F.~Bellini}
\author{L.~Cardani}
\altaffiliation{Also with: Physics Department, Princeton University, Princeton, NJ, USA}
\author{N.~Casali, I.~Colantoni, C.~Cosmelli, A.~Cruciani, I.~Dafinei, F.~Ferroni, S.~Morganti, P.J.~Mosteiro, F.~Orio, C.~Tomei, V.~Pettinacci, M.Vignati}
\affiliation{Dipartimento di Fisica, Universit\`a di Roma ``La Sapienza'' and INFN - Sezione di Roma, Roma, Italy}

\author{M.G.~Castellano}
\affiliation{IFN-CNR, Via Cineto Romano, Roma, Italy}

\author{C.~Nones}
\affiliation{Service de Physique des Particules, DSM/IRFU, CEA-Saclay, Saclay, France}

\author{T.D.~Gutierrez}
\affiliation{Physics Department, California Polytechnic State University, San Luis Obispo, CA, USA}

\author{X.G.~Cao, D.Q.~Fang, Y.G.~Ma, H.W.~Wang, X.G.~Deng}
\affiliation{Shanghai Institute of Applied Physics (SINAP), Shanghai, China}

\author{A.~Cazes, M.~De Jesus}
\affiliation{Institut de Physique Nucl\`eaire de Lyon, Universit\'e Claude Bernard, Lyon 1, Villeurbanne, France}

\author{B.~Margesin}
\affiliation{Fondazione Bruno Kessler, Trento, Italy }

\author{E.~Garcia}
\author{M.~Martinez}
\altaffiliation{Also with: Dipartimento di Fisica, Universit\`a di Roma ``La Sapienza'' and INFN - Sezione di Roma, Roma, Italy}
\author{J.~Puimedon, M.L.~Sarsa}
\affiliation{Universidad de Zaragoza, Laboratorio de Fisica Nuclear y Astroparticulas, Zaragoza, Spain}

\date{\today}

\begin{abstract} 
CUPID is a proposed future tonne-scale bolometric
neutrinoless double beta decay (\dbd) experiment 
to probe the Majorana nature of neutrinos and discover Lepton Number
Violation in the so-called inverted hierarchy region of the neutrino
mass. 
CUPID will be built on experience, expertise and
lessons learned in CUORE, and will exploit the current CUORE
infrastructure as much as possible. In order to achieve its ambitious
science goals, CUPID aims to increase the source mass and
dramatically reduce the backgrounds in the region of interest.
This requires isotopic enrichment, upgraded
purification and crystallization procedures, new detector
technologies, a stricter material selection, and possibly new shielding
concepts with respect to the state of the art deployed in CUORE. 
This document reviews and
describes the rich and varied R\&D activities, performed inside and
outside the CUORE collaboration towards the next generation
bolometric \dbd\ experiment.
  A separate document discusses the
science goals and timescale for CUPID in more detail.
\end{abstract}

\maketitle

\section{Introduction}\label{sec:intro}
This document, drafted by the CUPID Steering
Committee\footnote{CUPID steering committee: F.T. Avignone,
F. Bellini, C. Bucci, O. Cremonesi, F. Ferroni, A. Giuliani, P. Gorla,
K.M. Heeger, Yu.G. Kolomensky, M. Pallavicini, M. Pavan, S. Pirro,
M. Vignati}, describes a set of R\&D
activities -- performed in connection to the
present CUORE program -- which aim to develop technologies to be used
in a future bolometric neutrinoless double beta decay
($0 \nu \beta \beta$) experiment, introduced in
general in Ref.~\cite{CUPID-main}. CUPID is intended
to be a follow-up to the current CUORE experiment~\cite{CUORE}. It is
a proposed bolometric search which aims at a sensitivity to the
effective Majorana neutrino mass on the order of 10 meV, covering
entirely the so-called inverted hierarchy region of the neutrino mass
pattern. This primary objective poses a set of technical challenges:
the sensitive detector mass must be in the range of several hundred kg
to a ton of the isotope, and the background must be close to zero at
the ton~$\times$~year exposure scale. This objective can be achieved
by making use of the current CUORE infrastructure as much as
possible. However, four important upgrades with respect to CUORE have
to be taken into account:
\begin{enumerate} 
\item The increase of the sensitive mass demands the production of high-quality, radio-pure enriched crystals (see Section~\ref{sec:enr});
\item The detector technology needs to be upgraded with active
background rejection (see Sections~\ref{sec:strat},\ref{sec:R&D}); 
\item A higher radiopurity of materials is mandatory, together with related more sensitive trace analysis methods (see Section~\ref{sec:cont});
\item An active muon veto may be necessary (see Section~\ref{sec:veto}).
\end{enumerate}
All these topics require intense R\&D activities, which are described
below. While these activities are performed using resources separate from the CUORE
project, the current interest group believes that CUPID is the
proper framework for a possible final implementation of the experiment
based on technologies developed in this preliminary phase.  

\section{Source Production: Enrichment, Purification, Crystallization}\label{sec:enr}

\begin{figure}[b!]
\centering
\includegraphics[width=0.75\textwidth]{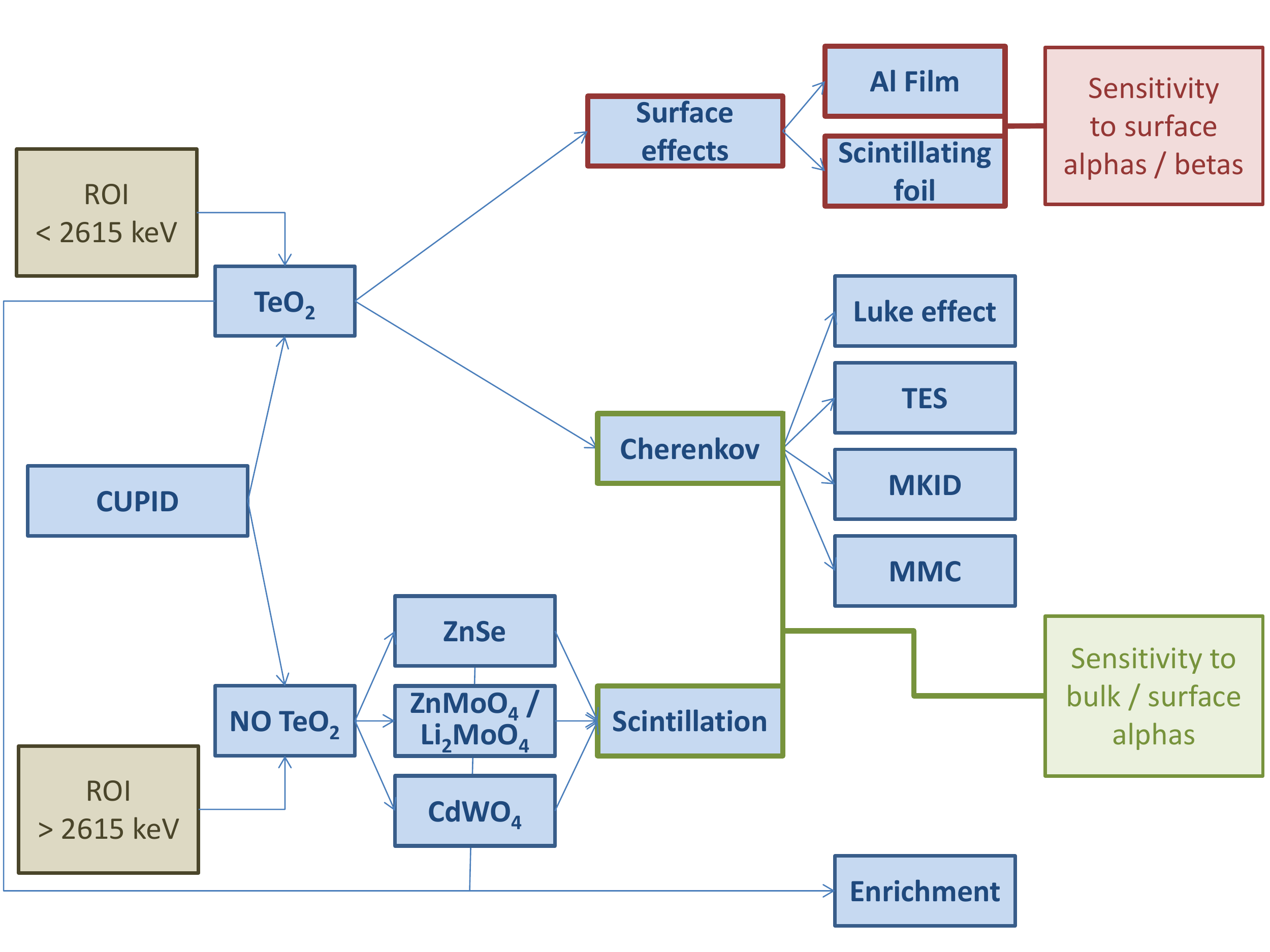}
\caption{Scheme of the R\&D detector activities for CUPID}
\label{fig:scheme}
\end{figure}

The efforts towards CUPID can be divided into two main categories:
those which focus on TeO$_2$ bolometers (and therefore the isotope
$^{130}$Te) and those which study alternative compounds, moving to
other isotopes  (see Fig.~\ref{fig:scheme}).  

CUPID will require isotopic enrichment for any isotope under
consideration. All of the isotopes discussed in the following --
$^{130}$Te, $^{100}$Mo, $^{82}$Se and $^{116}$Cd -- can be enriched by
centrifugation. This implies that they are all viable for a
next-generation experiment in terms of cost and production
rate. However, technical reasons determine differences in the
enrichment cost which may impact  the final choice. Very
approximately, the enrichment cost of $^{100}$Mo and $^{82}$Se is in
the range \$50-100/g. A factor of 2 more is expected for
$^{116}$Cd. The current enrichment cost of $^{130}$Te is \$17/g.  

Keeping TeO$_2$ would provide a big technological advantage in
terms of crystal growth, processing and radiopurity (all these issues
have already been successfully addressed in CUORE) and of isotope
enrichment as well. $^{130}$Te has a large natural isotopic abundance
(34 \%); as a consequence, the enrichment cost (for a final abundance $> 90$\%) is at least a factor 3 less than
for the other isotopes under consideration, as discussed
above. However, the use of isotopically enriched material changes
remarkably the purification / crystallization issues, for two main
reasons: first, the enriched material could have residual chemical
impurities which may demand additional purification stages to get
high-quality crystals; secondly, the enriched material is costly, and
therefore the growth procedure needs to be adapted in order to reduce
as much as possible the irrecoverable losses of the initial charge. An 
intense R\&D activity is ongoing in order to develop enriched TeO$_2$
crystals. The USC (University of South Carolina) group has procured about 10~kg of Te metal, enriched to 93\%
in $^{130}$Te. The first batch of CUORE-sized enriched 750\,g crystals
has recently been delivered to Gran Sasso for bolometric tests. The
initial performance is compatible with that of the unenriched CUORE
crystals. Further R\&D aiming at improving the efficiency of the
crystal growth process and reuse of the enriched material is ongoing.  

On the other hand, the level of background rejection required for the CUPID science
goals has already been demonstrated in alternative (scintillating)
crystals, albeit in single detector prototypes (see
Section~\ref{sec:noteo2}) and not yet in large arrays. Larger pilot experiments based 
on the isotopes $^{82}$Se (embedded in ZnSe crystals -- LUCIFER~\cite{Bee2013b} project) and $^{100}$Mo (embedded in ZnMoO$_4$ and in Li$_2$MoO$_4$
crystals -- LUMINEU~\cite{Ber2014} and LUCIFER projects) have been proposed and are under development. The candidate $^{116}$Cd (embedded in CdWO$_4$ crystals) is promising as well~\cite{cdwo4-bolo,Gironi-2009}. For
these isotopes, however, crystal production capabilities of several
manufacturers need to be investigated for a ton-scale
experiment. While the TeO$_2$ production line at the Shanghai
Institute of Ceramics of Chinese Academy of Science (SICCAS) has
performed extremely well for CUORE, large-scale production of other
crystals has never been attempted. Potential manufacturers in Ukraine and Russia have already been singled out, but investigating alternative vendors in the US, Europe, or Asia is desirable. The UCLA (University of California, Los Angeles) group plans to investigate establishing a ZnSe process at SICCAS, while other possible vendors are being considered.

R\&D activities towards the production of enriched crystals containing
$^{82}$Se, $^{100}$Mo, and $^{116}$Cd have already provided relevant
results. The procurement of a considerable amount of ultra-pure
$^{82}$Se by a European company (URENCO) represented a major
achievement by itself~\cite{Daf2014}. The enrichment is done by
SeF$_{6}$ centrifugation followed by the chemical conversion to
elemental selenium. In order to prevent radioactive contamination of
the samples, a dedicated centrifuge line and an ad-hoc conversion rig
were set up. 15 kilograms of Se, enriched to 95\% in $^{82}$Se, have
 been delivered. The overall chemical purity turns out to be better than
99.8\% on trace metal base; in particular, the concentrations of
$^{238}$U and $^{232}$Th fall below 10$^{-10}$g/g and the critical
impurities (Fe,Cr) have concentrations below the accepted limits for
good scintillation performance. The main effort is currently focused
on the refinement of the crystal growth procedure in terms of
optimization of the optical and thermal quality and limitation of the
irrecoverable loss of $^{82}$Se~\cite{Daf2014}.

Recently, two small ZnMoO$_4$ $^{100}$Mo  crystals enriched at 99.5\%
level (masses were on the order of 60 g) have been grown and
successfully operated as scintillating bolometers, with negligible
irrecoverable losses of the initial material ($<
4$\%)~\cite{Bar2014}. No deterioration of the bolometric and
scintillation performance of the enriched crystals was observed with
respect to the natural ones. Purification before crystal growth has
been done following a well-developed protocol which was used for large
mass natural crystals~\cite{Ber2014}. These results show that -- in
the case of $^{100}$Mo embedded in ZnMoO$_4$ crystals -- the
enrichment-purification-crystallization chain is well studied and
the achievements are close to those required by a large-scale
experiment.

In the case of CdWO$_4$ crystals for the study of the isotope
$^{116}$Cd, it is important to note that this compound (of course in
its natural form) is a well-established \emph{industrial} crystal
scintillator. Recently, a high-yield growth technology for enriched
$^{116}$CdWO$_4$ crystals was developed as well, with single crystal
masses up to several kg and with a production yield up to
90\%~\cite{Barabash-2011}. Within the ISOTTA~\cite{ISOTTA} project, 145~g of
99.85 \% enriched metallic $^{116}$Cd were recently (2014) purchased
to make dedicated crystallization and bolometric tests. The current
enrichment cost of \$240~/~g is high, although it can possibly be
reduced for mass-scale production.

\section{Strategies to Reduce the Background in the Region of Interest}\label{sec:strat}

In order to achieve the scientific goals mentioned in Section~\ref{sec:intro} and discussed in Ref.~\cite{CUPID-main},  the total background must be reduced by at least two orders of magnitude with respect to the current achievement. The objective is to attain a value on the order of 10$^{-1}$~counts/(ton y), demanded by the lifetime sensitivity target of $10^{27} - 10^{28}$ years. Marginal improvements in the energy resolution are possible, which is expected to be at best 1 keV FWHM for large crystals (the current value in CUORE-0~\cite{Q0}, which is expected also in CUORE~\cite{CUORE}, is 5 keV FWHM). Therefore, most of the efforts must be concentrated on reducing the number of background counts per unit energy around the ROI.

The expected dominant component of the background in CUORE is due to energy-degraded alpha particles emitted from the surfaces of the materials surrounding the detector or from the detector surface itself \cite{Arna08}. Given the enormous effort already devoted to surface treatment, it is not obvious that the required reductions in the background level can be achieved by improving the radiopurity of the detector materials alone (although this is an important R\&D goal). On the other hand, 
active background suppression promises the required levels, either with TeO$_2$ as sensitive material, or with other isotopes.  These ideas and the related activities performed so far are described in Section \ref{sec:R&D}. 

Improvement in the detector technology in discriminating detector-intrinsic background may not be sufficient. Background coming from residual environmental radioactivity and that induced by  sporadic muon interactions in the current CUORE configuration may produce backgrounds at the level $\sim 10^{-1}$~counts/(ton y). Discussions and mitigation strategies for these backgrounds are reported in Sections \ref{sec:cont}  and  \ref{sec:veto}, respectively. 

\section{Detector Technology R\&D}\label{sec:R&D}
 
The TeO$_2$-based R\&D activities aim at developing methods to identify and reject alpha particles in general (including therefore those related to surface contamination) or surface events in particular, by achieving sensitivity to the impact point.  One possible way to reject alpha particles is to detect the small amount of Cherenkov light emitted by the two electrons in the $0 \nu \beta \beta$ process, as alphas of the same energy are well below the Cherenkov threshold.  The essential component of the R\&D in this direction is to develop high-resolution light detectors capable of clearly identifying Cherenkov emission and therefore of separating alphas and betas on an event-by-event basis. Such light detectors require excellent resolution at the level of a few visible/UV photons. Several technologies are under study for the light detectors. Another approach aims at detecting surface events by achieving impact-point resolution, using either films deposited on the detector surface which can modify the signal shape of surface events, or surrounding the detector by a scintillating foil and detecting the consequent scintillation light emitted by a surface energy deposition.

The efforts on the alternative isotopes for CUPID, already introduced in Section~\ref{sec:enr}, are all based on scintillating bolometers. These devices use isotopes with a signal expected at around 3 MeV, and therefore above the last important gamma line of natural gamma background. The residual alpha background would be suppressed by exploiting the different light yield between alpha and beta particles, as in the aforementioned Cherenkov approach, but with at least an order of magnitude more light produced by scintillation. 

A graphical overview of the R\&D activities that are going on is reported in Fig.~\ref{fig:scheme}. Their present status is summarized in Table~\ref{tab:PID-status}

\begin{sidewaystable}\footnotesize
\caption{Status of the R\&D's on the various Particle Identification (PID) options explored in CUPID (see text for acronyms)}
\begin{tabular}{|x{1cm}|x{1.5cm}|x{3cm}|x{4cm}|x{2cm}|x{5cm}|x{5cm}|}
\hline
Isotope & Compound & Crystals & Enriched crystals & \multicolumn{2}{ c| } {PID technique} & PID status \tabularnewline
\hline
\multirow{7}{*}{$^{130}$Te } & \multirow{7}{*}{TeO$_2$ } & \multirow{7}{*}{\parbox{3cm}{\centering 125 cm$^3$ radio-pure natural crystals available from CUORE experience. Even larger natural crystals (216~cm$^3$) successfully grown.}} &  \multirow{7}{*}{\parbox{4cm}{\centering 54 cm$^3$ (125 cm$^3$) enriched crystals with a.i.=75\% (40\%) available and tested. R\&D in progress.}} & \multirow{2}{*}{\parbox{2cm}{\centering Surface effects}} & Al films & Bulk/surface rejection demonstrated with a 2 cm$^3$ crystal and a fast sensor  \tabularnewline \cline{6-7}
& & & & & Scintillating foil with conventional NTD light detector& Bulk/surface rejection partially achieved with a 125 cm$^3$ crystal \tabularnewline \cline{5-7}
& & & & \multirow{5}{*}{\parbox{2cm}{\centering Cherenkov light detection}} & Optimized NTD light detector & $\alpha$/$\beta$ rejection demonstrated with a 125 cm$^3$ crystal.  \tabularnewline \cline{6-7}
& & & & & NL-effect light detector & $\alpha$/$\beta$ rejection demonstrated with a 125 cm$^3$ crystal. \tabularnewline \cline{6-7}
& & & & & TES light detector & $\alpha$/$\beta$ rejection demonstrated with a 48 cm$^3$ crystal. \tabularnewline \cline{6-7}
& & & & & MKID & Prototype light detectors demonstrated. R\&D in progress.\tabularnewline \cline{6-7}
& & & & & MMC & Technology used in AMoRE. No test done on TeO$_2$. \tabularnewline \cline{6-7}
\hline
\multirow{2}{*}{$^{100}$Mo} & ZnMoO$_4$ & 78 cm$^3$ radio-pure natural crystals available. & 14 cm$^3$ enriched crystals with a.i.=99\% available and tested. 325 cm$^3$ crystalline boule with a.i.=99\% grown. R\&D in progress. &  \multirow{4}{*}{\parbox{2cm}{\centering Scintillation light detection }} & \multirow{4}{*}{\parbox{5cm}{\centering Conventional NTD light detector }} &  $\alpha$/$\beta$ rejection demonstrated with 78 cm$^3$ natural crystals and 14 cm$^3$ enriched crystals.  \tabularnewline \cline{2-4} \cline{7-7}
& Li$_2$MoO$_4$ & 78 cm$^3$ natural crystals available.& No test yet. & & & $\alpha$/$\beta$ rejection demonstrated with a 50 cm$^3$ natural crystal. 
 \tabularnewline \cline{1-4} \cline{7-7}
$^{82}$Se & ZnSe & 73 cm$^3$ radio-pure natural crystals available, but reproducibility problems. & No test yet. & & & $\alpha$/$\beta$ rejection demonstrated with a 73 cm$^3$ crystal but reproducibility problems.  \tabularnewline  \cline{1-4} \cline{7-7}
$^{116}$Cd & CdWO$_4$ & Large ($> 100$~cm$^3$) natural crystals available. & 75 cm$^3$ enriched crystals available. & & & $\alpha$/$\beta$ rejection demonstrated with a 63 cm$^3$ natural crystal.  \tabularnewline \hline

\end{tabular}
\label{tab:PID-status}
\end{sidewaystable}

\subsection{TeO$_2$ Bolometers}\label{sec:teo2}

As discussed above, the main expected limitation to the CUORE sensitivity arises from surface alpha contaminations \cite{Arna08}. To tag these background events in TeO$_2$ crystals two strategies are possible: (i) tag alpha particles through Cherenkov light or (ii) identify surface events. 

\subsubsection{Cherenkov Light Detection and Light Detector Technologies}

The threshold for Cherenkov emission in TeO$_2$ is around 50 keV for
electrons, and around 400 MeV for alphas~\cite{Tab2010}. The energy contained in the 
light produced by electrons above 350 nm, the transparency threshold at
room temperature, is computed to be 140 eV/MeV~\cite{Tab2010}. The
real value could be higher, since at low temperatures TeO$_2$ is
expected to become transparent at lower wavelengths. Experimental
tests confirm that light is not detected from alphas while, at the
$0 \nu \beta \beta$ energy, around 100 eV of light is detected from
beta/gamma particles~\cite{Cas2014}. Because of the high index of
refraction of TeO$_2$ ($n=2.4$), most of the light remains trapped in
the crystal, and the extraction of a higher light signal is difficult.
To reduce the alpha counts below the level of the beta/gamma
background predicted in CUORE, the signal to noise ratio in the light
detector must be greater than 5~\cite{Cas2014}. This implies that,
with a signal as small as 100 eV, the noise must be below 20 eV RMS, a
value that is difficult to reach. The light detectors used so far
consisted of germanium disks operated as bolometers, featuring an
average noise of 100 eV RMS. These detectors were developed by the
LUCIFER group to detect the high amount of scintillation light from
ZnSe and ZnMoO$_4$ crystals, but are clearly unsatisfactory to read
the tiny Cherenkov signal.  

As discussed below, separation on an event-by-event basis was recently achieved using a combination of Transition Edge Sensor (TES) readout and of Neganov-Luke (NL) effect~\cite{Wil2014}, and TES readout only~\cite{Sch2014}, for the light detector technology. Both results use  a large-area cryogenic light detector technology that already exists, and is based on TES developed in the framework of the CRESST Dark Matter experiment. This approach consists of
silicon-on-sapphire light absorbers read by a tungsten TES coupled to
an aluminum absorber.  This configuration can be used with CUORE-style bolometers, but scaling of the technology to a thousand
detectors requires extra R\&D. The most important issue is the
reproducibility of the technology (e.g. uniformity of transition
temperature $T_c$ across many channels) at temperatures of order 10 mK,
and the cost and effort required for the construction of a large
quantity of high-quality detectors. Additional aspects, such as
multiplexing of the detector signals to reduce the wiring complexity
and the heat load, would be useful; solutions already exist in the astrophysics community. 

Technologies under investigation now are: germanium bolometers implementing the NL effect,
different TES implementations, Microwave Kinetic Inductance Detector (MKID) sensors (CALDER project~\cite{CalderWeb}), and Magnetic Metallic Calorimeter (MMC) sensors. A carefully optimized Ge bolometer with a readout based on a simple neutron transmutation doped (NTD) Ge thermistor (as in the current CUORE technology) can also achieve the required performance~\cite{Cor2004}. A very recent still unpublished test has demonstrated a remarkable 2.9~$\sigma$ alpha/beta separation with a sophisticated NTD light detector and a CUORE-like TeO$_2$ crystal. 

It has to be stressed that the detection of Cherenkov light is
complementary to the use of scintillating foils (ABSuRD project~\cite{Canonica13}) or Al surface films to tag external betas or alphas~\cite{Non2012}. While with the
Cherenkov effect one can tag beta events and reject alphas and ``dark'' events
generated by lattice relaxations of the TeO$_2$ or by its supports,
with the other two methods one can identify beta and alpha background generated
outside or at the surface of the bolometer but not dark events. Low-noise light detectors
can also be used to read the scintillation light from ZnSe and CdWO$_4$ bolometers.
They could allow discriminating nuclear recoils from beta/gamma
interactions in the 10 keV region, enabling the search for Dark Matter
interactions in a way similar to that of the CRESST
experiment~\cite{Bee2013a}. On the other hand, molybdate crystals are not compatible with Dark Matter searches as they emit too low scintillation light.

\paragraph{Neganov-Luke effect \, --}
Light detectors with thresholds of a few eV can be developed by exploiting the so-called NL effect. In this approach, the light detector is an auxiliary bolometer consisting of a high purity Ge or Si wafer in which the ionization charges produced by the impinging Cherenkov photons are transported by an electric field. The work done by the field on the charges is detected as additional heat by a temperature sensor attached to the Ge/Si wafer.

Classical Ge bolometric detectors with NTD read-out have already been shown to be able to reach a threshold on the order of 50-100 eV for the total light energy in the LUCIFER context~\cite{Bee2013} and even better in carefully optimized devices to search for Dark Matter~\cite{Cor2004}. In parallel, amplification factors of one order of magnitude have been reached exploiting the NL effect, with constant noise. The combination of these two results makes it possible to achieve few-eV thresholds, in the regime of a few or even single optical - UV photon counting.  The quantum efficiency can also be very high, larger than 60\%, with a proper coating of the light absorber~\cite{Man2014}. 

In order to apply the electric field, different electrode designs will be compared. For detectors with Si absorbers, the contact pattern will be studied and fabricated by the Fondazione Bruno Kessler in collaboration with INFN Bicocca, which will take care of low temperature tests of the devices. For detectors with Ge absorbers, a structure of annular concentric Al contacts is foreseen, following the scheme adopted by the EDELWEISS collaboration for the charge read-out of their hybrid dark-matter bolometers. This configuration is well tested, also in terms of deposition procedure, and will be realized at CSNSM-Orsay in France, where the EDELWEISS detectors are usually produced. Prototype NL effect detectors based on this approach have been developed with encouraging results. Very recently, one of them has allowed achieving alpha/beta separation at a 2.6~$\sigma$ level in a CUORE-like TeO$_2$ crystal. This result is still unpublished.  As mentioned above, excellent preliminary results -- providing alpha/beta separation on an event-by-event basis -- have been achieved also with Si absorber, but using TES readout technology~\cite{Wil2014}. 

\paragraph{Transition Edge Sensor (TESs)  \, --}
Transition Edge Sensors are thin-film superconducting devices that
operate at the critical temperature $T_c$ of the superconductor. In
that transition region, TES devices have a large positive temperature
coefficient $\alpha$, which provides a sensitive measurement of
temperature. Changes in the TES current are detected by a
sensitive Superconducting Quantum Interference Device  (or SQUID array) amplifier, located at the still plate
(at $\sim 600$~mK) or at the 4~K plate of the dilution refrigerator. 

TES sensors have typically very low impedance, in the range of a few
m$\Omega$ to an $\Omega$. Therefore, they are inherently fast devices,
with a bandwidth of MHz or more. SQUID arrays can also provide
bandwidth in the MHz range. The large bandwidth compared to NTDs
offers several advantages: pulse shape sensitivity is significantly
improved, and time resolution better than 1 ms can be achieved,
reducing pileup due to $2\nu\beta\beta$ and background events. In
addition, the large bandwidth of the SQUID amplifiers allows relatively
straightforward time multiplexing of multiple sensors in a single
readout channel. Such solutions exist in the astrophysics community.

Very low-current noise of the SQUID amplifiers and the high
temperature coefficient of the TES sensors makes them very suitable
for high-resolution bolometric applications. Calculations show
potential for TES-SQUID based light detectors with eV-scale
resolutions, which has been demonstrated by the dark matter project
CRESST. In this experiment, W-TES achieved a $\sim 5$~eV RMS
resolution at zero energy. Typical T$_c$'s of W-TES operated in CRESST
are in the 15-20~mK range. As already mentioned, very promising
results -- providing alpha/beta separation on an event-by-event basis
in TeO$_2$ by detecting the Cherenkov light -- have been obtained with
the CRESST light-detector technology~\cite{Sch2014}. An interesting
alternative, which could allow reaching an even lower $T_c$, is a
thin-film bilayer of superconductor and normal conductor (e.g. Ir-Au,
Ir-Pt, Mo-Au, etc). The critical temperature depends then on the
stoichiometry or the ratio of thicknesses of the two films. UC
Berkeley and LBNL groups in collaboration with the Materials Sciences
group at Argonne National Laboratory are developing low-$T_c$ thin
films. Initial results are already encouraging, producing a sample
with $T_c$=21 mK and excellent temperature sensitivity. The next step
is developing high-resolution sensors based on these bilayers.

Another TES version under consideration as a sensor for the light detectors 
uses NbSi, which is a superconductor for an appropriate stoichiometric ratio with an intrinsic high resistivity in the normal state. The superconducting films will be fabricated at CSNSM, by ultra-high vacuum electron-beam co-evaporation of niobium and silicon on germanium substrates, according to a well established technology~\cite{Cra2011}. The films will have a meander structure obtained by reactive ion etching in order to further increase their normal state impedance up to the range 1-5~M$\Omega$. This high value allows the use of the same conventional electronics, based on Si JFETs, as for NTDs when they are operated within the transition. This solution does not provide all the advantages related to the low-impedance bilayer TESs (speed and ease of multiplexing), but it is possible to get a temperature sensitivity up to 10 times higher that that achieved by NTDs keeping the same front-end electronics, and so with a minimal impact on the CUORE readout structure. 
NbSi TESs with transitions around 20~mK have already been fabricated and tested. 

\paragraph{Microwave Kinetic Inductance Detectors (MKIDs)  \, --}
MKIDs base their working principle on the property of kinetic inductance in superconducting materials. The kinetic component of the impedance depends on the density of Cooper pairs, which can be modified by an energy release able to break them apart. If the inductive element is part of a resonant circuit with a high quality factor, the density variation of Cooper pairs generates changes in the transfer function of the circuit. The signal is obtained by exciting the circuit at the resonant frequency, and measuring the phase and amplitude variations induced by energy releases.

The main advantage of the KID technology resides in the ability to arrange parallel readout (multiplexing) and in the room temperature electronics, overcoming technical  issues related to the operation of TES. Many MKID sensors can be independently coupled to the same excitation line by making them resonate at slightly different frequencies. This feature allows using a small number of wires in the cryostat, simplifying the installation. The potential of MKIDs has been already demonstrated in astrophysical applications, where they successfully replaced the TES sensors.

The CALDER (Cryogenic wide-Area Light Detectors with Excellent Resolution) R\&D~\cite{Dido2014,CalderWeb}, supported by a European grant, aims at demonstrating the potential of  KID-based light detectors for CUPID. The first prototypes, based on aluminum sensors deposited on silicon substrates, are being tested as well as the readout. The first results are encouraging: both the detectors and the multiplexed readout work as expected. The CALDER group is now working on the choice of the superconducting materials and on the detector design to reach the noise goal.

\paragraph{Magnetic Metallic Calorimeters (MMCs)  \, --} 
MMC sensors are based on the strong temperature dependence of the magnetization in paramagnetic Au:Er sensors. 
A large variation of the magnetic moment  can be read out with high sensitivity using meander-shaped thin-film pickup coils and SQUID magnetometers. This effect, already exploited with outstanding results in X-ray spectroscopy by the Kirchhoff Institute for Physics (KIP) group in Heidelberg University, can be used to develop exceptionally sensitive thermometers the light detectors. The LUMINEU program, with which the Heidelberg group collaborates, envisages the fabrication of devices based on this principle. The KIP group will develop new Au:Er sensors with specific meander geometry for optimized meander inductance and sensor heat capacity for LUMINEU~\cite{Loi2014}. An innovative deposition of the meander directly on the Ge crystal will be developed to ensure the very fast readout of the entire wafer with only one sensor. Signal shape, amplitude and noise calculations, usually very reliable in MMCs, foresee an energy resolution between 3 eV and 10 eV (FWHM) and the signal rise-time below 50~$\mu$s. Encouraging tests with actual devices have already been performed.
 
\subsubsection{Al Thin Film as Signal Shape Modifiers}
Tagging surface events is difficult, as -- due to their basic working principle -- bolometers do not have a dead layer (they are fully sensitive up to the surface) and often present a single response to any type of fast energy deposition, irrespective of its nature and location (e.g. in the CUORE TeO$_2$ case). While this property is responsible for excellent energy resolution and detection efficiency with little position dependence, 
it can be unfortunate when surface events are the dominant background, since surface is as sensitive as the bulk. This disadvantage can be overcome by adding passive elements to the bolometer surface in a reproducible way and with radio-clean procedures.  Surface sensitivity can be achieved by depositing Al films (of $\mathcal{O}(10~\mu\mathrm{m})$ thickness) on the main bolometric TeO$_2$ absorber. 

The rationale of this approach is the following. Athermal phonons generated by a particle that releases its energy within a few mm from the surface (i.e. an alpha or beta particle) will break Cooper pairs in the superconducting film and produce quasiparticles, which have in general a long lifetime (on the order of milliseconds) in high purity aluminum. Their subsequent recombination will produce athermal phonons again, that will add a delayed component to the phonon signal read out by the sensor on the main bolometric absorber. We expect remarkable difference in signal formation for bulk events. In this case, the athermal phonon population reaching the Al film is more degraded in energy and less efficient in producing quasiparticles. Clear evidence of this mechanism has already been achieved~\cite{Oli2008}. Consequently one expects different signal shapes for surface and bulk events, and in particular a longer rise-time for a surface event.  

The proof of principle of this approach was already demonstrated with TeO$_2$ in CSNSM-Orsay~\cite{Non2012}, but using fast phonon sensors based on NbSi films, with rise times on the order of 1 ms. Not only the rise-time was longer for surface events, but the pulse shape was modified for several milliseconds after the maximum. Unfortunately, the current NbSi sensor technology does not provide an adequate energy resolution, and is therefore unsuitable for $0 \nu \beta \beta$. The future R\&D work, to be performed at CSNSM-Orsay and CEA/SPP-Saclay, will consist of achieving surface-to-bulk signal separation by pulse shape discrimination with NTD sensors, i.e. with a heat pulse rise-time on the order of tens of milliseconds. This may be possible as the excellent signal-to-noise ratio characterizing the typical CUORE readout has the potential to highlight even tiny pulse-shape differences. 
%
%

Alternatively, low-impedance bilayer TES sensors may be more suitable for pulse shape discrimination, due to their inherent fast response time and excellent noise characteristics. Once such sensors are developed, we will study their applications in pulse shape discrimination. This work is proceeding at Berkeley. 

Once pulse shape discrimination is demonstrated above ground in a small prototype, a procedure will be set up to deposit  Al films on all six sides of a typical $5 \times 5 \times 5$~cm TeO$_2$ crystal, with the aim to proceed to underground tests on real size detectors. It is noted that this technology does not involve light detectors and their additional readout.

\subsubsection{Surface Event Detection Mediated by a Scintillating Foil}
Another way to tag surface events is proposed in the ABSuRD project (A Background SUrface Rejection Detector). In this approach, surface energy depositions can be identified by means of an external plastic scintillator. The idea is to encapsulate a purely thermal bolometer (such as TeO$_2$) with a scintillating foil and to add a bolometric light detector to measure the light. When degraded alpha particles (or surface-originated beta particles) interact in the scintillating foil, the emitted light is collected by the light detector. A surface alpha particle releasing part of its energy in the crystal and part in the scintillating foil can be rejected by analyzing the coincidence signal of heat (in the absorber) and light (in the light detector) \cite{BGS}. The crucial aspect of this technique is the capability of detecting the light emitted by the scintillating foil. A moderately low energy threshold of $\sim$ 1 keV is needed for the bolometric light detector. As an example, an alpha particle of 5.3 MeV (generated by the decay of $^{210}$Po) releasing 2.5 MeV in the scintillator (i.e. generating a 2.8 MeV background event in  the TeO$_2$ bolometer) produces about 1.5 - 2 keV of photons. A clear advantage of this technique is that beta radiation escaping the detector can be tagged as well due to the high scintillation of electrons, in spite of the small deposited energy. The ABSuRD project is focused on developing and characterizing plastic scintillators with large light yield and good low temperature properties to minimize the impact of the light detector energy threshold. A first bolometric prototype, realized with commercial scintillators, showed encouraging results in the capability of tagging surface alphas from an implanted $^{147}$Sm source \cite{Canonica13}. 

\subsection{Alternatives to TeO$_2$ Bolometers}\label{sec:noteo2}
For bolometric applications, the most promising $0 \nu \beta \beta$ isotopes alternative to $^{130}$Te are $^{82}$Se, $^{100}$Mo, and $^{116}$Cd. They are all characterized by a $0 \nu \beta \beta$ transition energy higher than 2.6 MeV, in a region free from gamma background. The residual and presumably dominant alpha background will be rejected by embedding the isotopes under study in scintillating bolometers. The different light-to-heat ratio for alpha and gamma/beta interactions for the same thermal energy is a powerful tool to identify and eliminate alpha events.

\subsubsection{Study of $^{82}$Se Embedded in ZnSe Crystals}
ZnSe crystals represent a very interesting candidate for the search for the $0 \nu \beta \beta$ of $^{82}$Se by virtue of its high content of Se (56\%) and high Q-value (Q$_{\beta\beta}$ = 2997 keV), as well as its good bolometric and scintillating properties. Several Zn$^{nat}$Se bolometers were studied in the framework of the LUCIFER project, supported by a European grant. In particular a 430\,g   Zn$^{nat}$Se crystal was characterized in terms of the energy resolution, internal contaminations, and particle identification capabilities~\cite{Bee2013a}. The crystal exhibited a FWHM energy resolution of 16.3 keV at 2615 keV and a light yield of ~6.5~keV/MeV and 27~keV/MeV for beta/gamma and alpha particles respectively.  The possibility to discriminate alpha events from beta/gamma interactions by pulse shape due to the difference in the decay constant of the scintillation pulses was demonstrated. An alpha discrimination power $>$99.99\% was achieved combining the pulse shape and light yield discrimination. 

A contamination of $\sim$17 $\mu$Bq/kg in $^{232}$Th and of $\sim$25$\mu$Bq/kg in $^{238}$U in secular equilibrium was measured in a dedicated 524-hour underground run. These contaminations, obtained without particular care in material selection, are compatible with a background level of 1~count/(keV ton y) in the region of interest. 

The LUCIFER light detector~\cite{Bee2013b} is a disk-shaped pure Ge bolometer (\O = 44~mm, thickness = 180~$\mu$m) with a SiO$_2$ dark coating~\cite{Bee2013d} on the side facing the main scintillating crystal. The light detector is read out with an NTD thermistor. FWHM energy resolution of ~220 eV and rise time of a few ms were typically observed~\cite{Bee2013c}. These performances are suitable for particle discrimination in ZnSe bolometers given the light yield and the time profile of the scintillating pulse. 

As reported in Section~\ref{sec:enr}, enrichment, purification and crystallization are an important topic of the LUCIFER program. The results in terms of production of pure isotope are excellent, whereas the growth procedure is still under refinement~\cite{Daf2014}.  

LUCIFER aims at running a pilot experiment with a total isotope content of 10 -- 15 kg and a background level of $10^{-3}$~count/(keV kg y) in order to demonstrate the maturity of the technology~\cite{Bee2013b}.  The LUCIFER detector array will be installed in the dilution cryostat presently hosting CUORE-0. Data taking is foreseen in Spring 2016.

\subsubsection{Study of $^{100}$Mo Embedded in ZnMoO$_4$ or Li$_2$MoO$_4$ Crystals}
$^{100}$Mo is one of the most promising $0 \nu \beta \beta$ isotopes because of its high transition energy (3034 keV), outside the bulk of the natural gamma radioactivity, and its considerable natural isotopic abundance (9.7\%). The best sensitivity to $0 \nu \beta \beta$ of $^{100}$Mo was reached by the NEMO-3 experiment, which obtained a half-life limit of $1.1 \times 10^{24}$~y at 90\% C.L. with $\sim$7 kg of enriched $^{100}$Mo and 4.5 y live time \cite{Arn14}. The NEMO-3 detection efficiency (14\%) and energy resolution (10\%) can be improved up to 80-90\% and to $\sim$0.1\% respectively by using bolometers containing Mo. Once again, if Mo is embedded in a scintillating crystal it is possible to develop hybrid devices with a double heat+light readout aiming at a full suppression of the alpha background. As discussed above, the most natural and effective device to detect scintillation photons is a dedicated bolometer.
 
There are several inorganic scintillators containing Mo. One of the most convenient choices consists of ZnMoO$_4$ crystals. Recent developments show that this material is very promising for a high sensitivity $0 \nu \beta \beta$ experiment \cite{Bee2013b,Bee2012a,Bee2012b,Bee2012c}. Energy resolutions better than 10 keV FWHM have been routinely obtained with crystals up to 330\,g mass in energy regions close to the $0 \nu \beta \beta$ signal position, in the framework of the LUCIFER, LUMINEU, and ISOTTA projects. Light yields on the order of 1 keV/MeV, moderate but sufficient, were demonstrated. Underground tests of these large mass crystals, both in the Gran Sasso and in the Modane laboratories, have shown that $^{228}$Th -- the most critical contaminant due to emission of high-energy beta particles -- has a specific activity less than 5 $\mu$Bq/kg~\cite{Bee2012c}. In addition, alpha discrimination power much better than 99.9\% has been demonstrated by comparing alpha and beta/gamma light yield. This figure is compatible with a background level in the range of a few $10^{-1}$~counts/(ton y)~\cite{Bee2012a,CUORE-IHE}.  
Differences in pulse shapes between alpha- and gamma-induced heat signals has been observed, suggesting an even better rejection of alpha events using pulse shape discrimination alone~\cite{Bee2012c}. This aspect deserves to be studied further, as it can potentially lead to eliminating the need for a light sensor, which would simplify the detector structure. 

An additional potential background source in $^{100}$Mo-based bolometers is due to random coincidences of the standard two-neutrino double beta decay ($2 \nu \beta \beta$) events~\cite{Bee2012a,Che2012}. This process has recently been observed in a bolometric experiment~\cite{Car2014}. 
The relatively high rate ($T_{1/2} = 7.11 \times 10^{18}$~y) of $2 \nu \beta \beta$ events makes the pileup problem particularly acute for $^{100}$Mo. Time resolution of significantly better than 1\,ms is required to reduce this background to a negligible level~\cite{CUORE-IHE}. Pileup events can also be rejected using pulse-shape discrimination~\cite{Che2012,Che2014}. 

The production of medium-volume radio-pure enriched crystals with excellent bolometric performance (see Section~\ref{sec:enr}) is an achieved result in the LUMINEU program~\cite{Bar2014}. The only possible showstopper could be the remarkable difficulty encountered in growing large-mass regular-shape crystals. However, recent crystallization tests at the Nikolaev Institute of Inorganic Chemistry (Novosibirsk, Russia) exhibit  significant improvements~\cite{Ber2014, Che2013}, culminating in the production of high-quality cylindrical samples (h=4~cm and \O=5~cm). 

The next step for $^{100}$Mo developments will involve a medium-scale pilot experiment, to be performed on a two-year time scale at LNGS and/or in the Modane underground laboratory. 
Such a detector will consist of tens of $\sim$400 g ZnMoO$_4$ enriched crystals using approximately 10 kg of the isotope, belonging to ITEP (Russia) and already made available for this search. This demonstrator, with a remarkable sensitivity by itself, will constitute a general feasibility test for the use of this technology at the ton-scale level. A Memorandum of Understanding among IN2P3 (France), INFN (Italy), and ITEP (Russia) has been signed with respect to this technology demonstrator, guaranteeing in particular the use of the existing enriched $^{100}$Mo for this development.

An alternative compound to ZnMoO$_4$ is Li$_2$MoO$_4$, which is encouraging in spite of a low light yield (on the order of $\sim 0.6$ keV/MeV, about half of what has been observed in ZnMoO$_4$), as preliminary tests~\cite{Car2013,Bek2014} in the framework of ISOTTA have shown. The main advantage of this compound is that, according to all the preliminary indications, the synthesis of large mass crystals is much easier than for ZnMoO$_4$. Li$_2$MoO$_4$ has been used as a solvent for the growth of many refractory oxides in the past, and indeed it was investigated as such. Consequently, many of its physical and physicochemical properties are known, including their temperature dependencies, which makes this system amenable to advanced numerical simulations of the growth process aiming at up-scaling the latter for the production foreseen in a one-ton experiment. Li$_2$MoO$_4$ is congruently melting at ~705$^{\circ}$C, that is about 320$^{\circ}$C below the melting point of ZnMoO$_4$ crystals, which entails a virtually nil volatilization. In ZnMoO$_4$ crystal growth, on the contrary, such volatilization phenomena may induce $^{100}$Mo losses as well as parasitic inter-growths in the first stages of growth which, in addition to a very low crystal symmetry, lead to a crystal shape that is difficult to control. In Li$_2$MoO$_4$ crystals, the highest symmetry permits to obtain very easily the growth orientation and to control it by proper seed preparation, as well as constant diameter and regular crystals. Moreover, the lower melting entropy of Li$_2$MoO$_4$ crystals with respect to that of ZnMoO$_4$ crystals also avoids in Li$_2$MoO$_4$the annoying faceting of the ZnMoO$_4$ crystals, which contributes to their irregular shapes. In the perspective of an industrial production, to work at lower temperatures also decreases the contamination kinetics of the Pt crucibles to be used, an so enhances their life cycle. Last but not least, another advantage of Li$_2$MoO$_4$ crystals is their highest Debye temperature associated with zero-point energy, as compared with that of ZnMoO$_4$ crystals.

A test performed above ground at CSNSM in late 2014 with a relatively large volume Li$_2$MoO$_4$ crystal (h=4~cm and \O=4~cm) has shown an excellent bolometric perfomance and alpha/beta discrimination factor, at the level of or even better than those achieved in ZnMoO$_4$~\cite{Bek2014}. The same detector operated underground in LNGS has confirmed and improved these results, and has shown also a remarkable radiopurity of this compound.  In the first part of 2015, an intense activity on Li$_2$MoO$_4$ -- involving also enriched crystals -- will clarify if it may validly replace ZnMoO$_4$ in the $\sim 10$ kg isotope experiment described in the previous paragraph. 

\subsubsection{Study of $^{116}$Cd Embedded in CdWO$_4$ Crystals}

As $^{100}$Mo and $^{82}$Se, $^{116}$Cd belongs to the  so-called ``$0 \nu \beta \beta$ golden isotopes", i.e. elements whose $0 \nu \beta \beta$ transition energy exceeds the 2615 keV $\gamma$-line of $^{208}$Tl. Enriched $^{116}$CdWO$_4$ scintillating bolometers could be ideal candidates for a $0 \nu \beta \beta$ searches for several reasons, in part already discussed in Section~\ref{sec:enr} as far as enrichment and radio-pure crystallization are concerned~\cite{Barabash-2011}. In addition, the light yield is comparable to the best known undoped scintillators and the radiopurity of this compound is ``naturally'' high ~\cite{Danevich-2003}.

Due to these favorable features this enriched crystal compound was already used to perform a $0 \nu \beta \beta$ experiment~\cite{Danevich-2003, Poda2014} using 
standard photomultipliers as light sensors.
Several tests ~\cite{cdwo4-bolo,Gironi-2009} were also performed on scintillating CdWO$_4$ bolometers, demonstrating the  
high energy resolution, the low level of contaminations, and the excellent particle discrimination capabilities of such detectors.
Moreover, thanks to the extremely high light yield, these crystals can be potential candidates for dark matter searches.
In fact, the light yield as well as the intrinsic radiopurity of this compound are better than the CaW$O_4$ crystals used in the CRESST experiment.

The only drawback of this compound is the presence, in the natural Cd, of the $^{113}$Cd isotope (12\% i.a.) that undergoes $\beta$-decay with a Q-value of 319~keV and  $T_{1/2}$ = 9.3 $\times$ 10$^{15}$ y (roughly 0.5~Hz for 1~kg of CdWO$_4$). 
This  results in an unavoidable pile-up background spectrum at the Q$_{\beta\beta}$ value due to the convolution with the 2$\nu$-DBD. In addition, the low-energy $^{113}$Cd $\beta$-decay prevents -- definitively -- the use of this compound for Dark Matter search. In the future bolometric $0 \nu \beta \beta$ experiment, in which isotopic enrichment is mandatory, these negative effects will be automatically overcome. For example, the measured level of  $^{113}$Cd in the sample produced in the framework of ISOTTA (see Section~\ref{sec:enr}) is $\leq$0.02 \%, which is adequate for both $0 \nu \beta \beta$ and dark matter searches. 

\section{Reduction of the Environmental Radioactivity} \label{sec:cont}
The active background rejection techniques, on which detector developments are focused, aim at reducing to negligible levels the effect of surface contaminations of detector materials. As mentioned, this background source is identified as the dominant contributor to Cuoricino and CUORE-0 counting rates and as the most likely limiting factor for CUORE sensitivity. However, the reduction of surface contamination effects can't by itself ensure the achievement of a background level two orders of magnitude lower than CUORE. Indeed, sources different from surface contaminations can contribute to the ROI counting rate at levels below the 10~counts/(keV ton y) foreseen for CUORE. Among these, the most dangerous are certainly the radioactive (bulk) contaminations of the detector elements: crystals, copper, lead, and the ``small parts'' as glue, heaters, bonding wires or flat cables and pads etc.
In the CUORE background budget~\cite{CUPID-main,CuoreBB}, no positive indication of the presence of contaminants in the different detector elements have been obtained, with the exception of the surface contaminations. However, in most cases, the available upper limits on material contamination translate to potentially dangerous counting rates for the CUPID goal. For this reason, an improvement of the presently attained sensitivities is mandatory.

Two techniques  will be explored in the CUPID program:
\begin{itemize}
\item Pre-concentration of radio-contaminants through chemical treatment of materials. When coupled to NAA, ICPMS, or HPGe screening this will allow increasing the sensitivities achieved with these techniques. As an example, in the case of copper either NAA and HPGe spectroscopy achieve a similar sensitivity of about 1 $\mu$Bq/kg on $^{232}$Th. In both cases the sensitivity is ultimately limited by the mass of the copper sample that can't be increased \emph{ad libitum} (increasing the mass of a sample measured with HPGe is often not effective due to the self-absorption of the gamma lines inside the sample). A pre-concentration of the contaminant is equivalent to an increase in the mass of the sample resulting therefore in a sensitivity increase. The technique is often used in ICPMS measurements but can also be successfully applied either in NAA or in HPGe measurements. However, it requires a dedicated study for each material, where the achievement of the required concentration
  as well as the control of systematics has to be proven. 
\item Development of a bolometric detector for the measurement of surface/bulk contamination of small samples and foils. In a number of cases the aforementioned screening techniques can't be applied either because the available mass sample is too small (HPGe spectroscopy require large mass samples to reach high sensitivity) or because the material properties are inappropriate (NAA and ICPMS have restrictive conditions on the chemical elements that can be analyzed). For CUORE this was true in the case of ``small parts'' or of materials used in the form of foils (superinsulation, flat cables, etc). In these cases, the use of surface alpha spectroscopy through Si surface barrier diodes has been proven to reach competitive sensitivities in shorter times. The same kind of screening, implemented through the use of bolometric detectors, will achieve a sensitivity that is between 10 and 100 times higher thanks to the better energy resolution of these devices as well as to their higher radiopurity. Si wafers or TeO$_2$ slabs can be  used for this purpose realizing a sandwich-like detector where samples are inserted in-between thin bolometers. If able to reach very low thresholds these detectors could also provide information on the X-ray emission of the samples providing in this way complementary information for contamination identification.
\end{itemize} 
If successful, the improvement reached with the development of these two technologies will allow us to reach a sensitivity on U and Th concentrations of about one to two orders of magnitude higher than what is obtained today, which is the required goal for CUPID materials.

\section{Control of Muon-Induced Radioactivity} \label{sec:veto}

According to the simulations~\cite{CUORE-IHE} based on CUORE and the measured muon flux at LNGS, the event rate induced by the cosmic ray muons or muon showers in the $0\nu\beta\beta$ region of interest is expected to be on the order of $0.5$~counts/(ton y). For a 5-10 year exposure, a reduction in this rate of about a factor of 10 or more would be required for a zero-background experiment. Such reduction would require either a deeper underground site, or a dedicated anti-muon veto around the active volume of the CUPID detector. The muon-induced neutron rate is estimated to be another order of magnitude smaller; however, simulating muon-induced showers with high precision is a notoriously difficult endeavor. 

The Yale group of the CUORE collaboration is developing a muon tagger around CUORE as part of the R\&D towards a future CUPID experiment. The goals are to enable a data-driven study of muon-induced backgrounds in CUORE and to confirm that LNGS is deep enough for a future bolometric experiment like CUPID.

\section{Conclusions}

Ton-scale bolometric detectors based on an upgraded CUORE technology and infrastructure have the potential to convincingly discover or rule out the Majorana nature of neutrinos in the so-called inverted neutrino mass hierarchy~\cite{CUORE-IHE}. This document describes the R\&D efforts towards the next-generation bolometric experiment CUPID. The objective of this future project is to achieve at least a $3 \sigma$ discovery potential for $0\nu\beta\beta$ decay if the neutrino is a Majorana particle and if the mass hierarchy is inverted. 

The experimental activities reviewed in this document focus on a background reduction beyond that foreseen in CUORE and demonstrated by CUORE-0. They are currently performed either within the CUORE collaboration, or as independent projects for small-scale $0\nu\beta\beta$ demonstrators. These activities span a wide range of approaches, methods, technologies and materials, involving tens of physicists even outside the current CUORE collaboration. This testifies to the strong scientific attraction exerted by a next-generation bolometric $0\nu\beta\beta$ experiment. This richness is a positive factor in the present phase, but it is important to converge towards a well-defined structure for CUPID in a two-year time scale, in order that this experiment be timely in the global context of $0\nu\beta\beta$ decay. This convergence process is discussed in Ref.~\cite{CUPID-main}. 



%
%
%

\end{document}